# Tailoring Mechanical Properties of Germanium Anodes via Metal Incorporation for Improved Cycle Stability

Koki Nozawa[1,a)], Noriyuki Saitoh[2], Noriko Yoshizawa[3], Takashi Suemasu[1], and Kaoru Toko[1,a)]

[1] Institute of Applied Physics, University of Tsukuba, 1-1-1 Tennodai, Tsukuba, Ibaraki 305-8573, Japan

[2] Electron Microscope Facility, AIST, 1-1-1 Higashi, Tsukuba, Ibaraki 305-8564, Japan

[3] Department of Energy and Environment Research Planning Office of Zero Emission, AIST, 16-1 Onogawa, Tsukuba, Ibaraki 305-8569, Japan

[a)] Authors to whom correspondence should be addressed: nozawa.koki.td@alumni.tsukuba.ac.jp; toko@bk.tsukuba.ac.jp

Achieving long-term stability in high-capacity lithium-ion battery anodes remains a critical challenge. In this study, we present a materials-intrinsic strategy for extending the cycle life of Ge, a promising next-generation anode material, through trace doping with metal elements. We systematically investigated the effects of small additions of various metals and found that elements with large atomic size, particularly Yb, markedly improved the cycling stability without sacrificing the initial capacity, while appropriate Yb doping enhanced the anode lifetime by approximately a factor of three. Structural and electrochemical analyses revealed that this improvement originates from mechanical softening of the Ge anode, which suppresses lithiation-induced damage such as cracking and delamination. Nanoindentation measurements further showed a strong negative correlation between dopant atomic size and film hardness, establishing anode softening as a new design principle for damage-tolerant electrodes. Although Yb doping reduced the rate capability at high C-rates, the present results demonstrate a clear shift in design strategy from volume-change suppression to mechanical compliance. These findings provide a useful framework for stabilizing high-capacity alloy anodes through atomic-scale mechanical control.

## 1. Introduction

Lithium-ion batteries (LIBs) are the dominant energy storage technology for consumer electronics, electric vehicles, and grid applications, owing to their high energy density, long cycle life, and operational reliability [1,2]. However, the growing demand for higher energy and power densities, especially in electric vehicles and grid-scale storage, is pushing conventional LIBs to their theoretical limits. One of the main bottlenecks lies in the limited capacity of graphite anodes, which store Li ions through intercalation, with a theoretical maximum capacity of 372 mAh $g^{-1}$ [3−5]. This intrinsic limitation has motivated the exploration of alternative anode materials that can provide substantially higher Li storage capacities.

The most promising alternatives include group IV elements such as Si, Ge, and Sn, which form Li-rich alloys during battery operation [6−9]. These materials exhibit much higher theoretical capacities, that is, approximately 3579 mAh $g^{-1}$ for Si, 1600 mAh $g^{-1}$ for Ge, and 994 mAh $g^{-1}$ for Sn, making them attractive candidates for next-generation high-energy-density batteries [6]. In particular, Ge has emerged as a compelling option due to several key advantages over its counterparts. Ge has a Li diffusivity approximately 400 times higher than Si and an electrical conductivity that is four orders of magnitude greater [10−13]. These properties enable faster kinetics and improved rate capability, making Ge suitable for high-power and fast-charging applications.

Despite these benefits, the application of Ge anodes is hindered by significant mechanical degradation during repeated lithiation and delithiation [14,15]. The large volume expansion of Ge (~330%) leads to internal stress, cracking, pulverization and delamination from the current collector, all of which result in capacity fading and short cycle life. Although the volume change is smaller than that of Si (~420%), it is still substantial enough to cause electrode failure over extended cycling [16]. Moreover, the degradation of Ge anodes is increasingly recognized as a coupled electrochemical–mechanical process, in which stress-driven cracking and delamination expose fresh surfaces for repeated SEI re-formation and locally perturb Li-ion transport, thereby accelerating fracture propagation and capacity decay during cycling[17]. Previous strategies to address these issues, such as, nanostructuring [18−21], composite formation [22−27], amorphization [28,29], defect engineering[30,31], and advanced binder systems [32−38], have demonstrated partial success but often involve complex synthesis routes or trade-offs in practical energy density and manufacturability.

One emerging approach involves incorporating electrochemically inactive metal elements into the anode to reduce volume expansion during cycling [39−45]. However, the inclusion of these elements inevitably reduces the initial capacity of the anode. As a result, although moderate improvements in cycle life have been reported, the substantial loss in capacity often outweighs the benefits, limiting the practical impact of this strategy. To date, this trade-off between mechanical stabilization and capacity reduction has posed a major challenge, and a clear pathway to simultaneously achieving both durability and high capacity remains elusive.

Here, we present a materials design strategy to extend the cycle life of Ge-based anodes by leveraging mechanical softening via trace metal incorporation with the aim of elucidating the underlying mechanisms responsible for metal-addition-induced Ge anode property enhancement. We show that the incorporation of small amounts of various metal elements into Ge, particularly those with large atomic radii such as Yb, markedly improves cycling performance. Importantly, because the incorporation level is kept low, the initial capacity remains virtually unchanged, thereby avoiding the trade-off typically associated with the addition of electrochemically inactive elements. Nanoindentation analysis further reveals a strong negative correlation between the atomic radius of the incorporated element and the film hardness, suggesting that the resulting softening of the anode enhances mechanical compliance and damage tolerance during cycling without sacrificing initial capacity.

## 2. Experiment

Metal (M = Al, Cu, Ni, Ag, W, Ta, and Yb) chips were sequentially cleaned with acetone, methanol, and distilled water, and subsequently affixed to the Ge sputtering target using carbon adhesive tapes. The purity of the Ge sputtering target was 99.9%. The purity and thickness of the metal chips were 99.9% and 50 μm, respectively. $Ge_{1-x}M_x$ thin films (500 nm thickness) were prepared on Mo foils using radio-frequency (RF) magnetron sputtering (Sanyu Electron SVC-700RF, base pressure: $3.0 \times 10^{-4}$ Pa) at room temperature with Ar plasma. The RF power was set to 100 W, yielding a deposition rate of 20 nm $min^{-1}$. The composition $x$ was controlled by adjusting the placement and number of metal chips on the target. Out-of-plane ($\theta$–$2\theta$) X-ray diffraction (XRD) patterns were measured using a Rigaku SmartLab system with a Ge monochromator (wavelength: 1.54 Å) and a Cu Kα radiation source (voltage: 40 kV, current: 30 mA). Raman spectroscopy was conducted using a JASCO NRS-5100 (laser wavelength: 532 nm; spot size: 5 μm), and the laser power (0.5 mW). The samples were punched through a 10 mm diameter disk and then used as anodes after drying at 120 °C for 12 h in vacuum. In an Ar-filled glove box (UNICO UL-Ef1000A-TKSP), coin-type cells were fabricated from a $Ge_{1-x}M_x$ anode, pure Li metal foil, and a separator (Celgard 2400) immersed in an electrolyte. The electrolyte was Li hexafluorophosphate (1 mol $L^{-1}$) in ethylene carbonate/diethyl carbonate (1:1 in volume). The electrochemical characteristics of the samples were investigated using a multichannel galvanostatic potentiostat (Bio-Logic VMP). For each composition, electrochemical measurements were performed on three independently fabricated cells. For clarity of presentation, the best-performing cell is shown as a representative result. The variation among cells was small, and the observed differences in electrochemical performance fell within the scatter of the plotted data. The cells were disassembled using a coin-type cell disassembly machine (Hohsen). The samples were then retrieved and cleaned with distilled water. For the evaluation of solid electrolyte interphase (SEI), X-ray photoelectron spectroscopy (XPS) measurements were carried out using an XPS system (JEOL JPS-9010TR). The $Ge_{1-x}M_x$ morphology was evaluated using a SEM (Hitachi High-Technologies SU-7000) with an energy dispersive x-ray spectroscopy (EDX) (Oxford AZtec). An FEI Tecnai Osiris instrument operating at 200 kV was used for TEM analysis. Cross-sections of the samples were prepared using an focused-ion beam (FIB) (Helios Nanolab 600i). Nanoindentation tests were performed using a Nano Indenter (Nano indenter-XP) via the continuous stiffness measurement method. The tip area function was calibrated using a standard fused-silica sample prior to the measurements. Indentations were performed on visually smooth regions of films deposited on flat glass/$SiO_2$ substrates, while avoiding defects and debris.

The $Ge_{0.97}M_{0.03}$ model was constructed using a supercell approach. A 4 × 2 × 2 supercell containing 32 Ge atoms was generated from the primitive Ge unit cell, and one Ge atom out of the 32 atoms was substituted by a metal atom, corresponding to a metal concentration of 3.125% (≈3%). A 4 × 2 × 2 supercell containing 32 atoms was generated from the primitive Ge unit cell. This substitutional model was adopted to represent metal incorporation in the amorphous Ge matrix in a simplified manner suitable for comparative analysis of mechanical properties. First-principles calculations were performed using the Vienna Ab initio Simulation Package (VASP) [46,47]. A plane-wave cutoff energy of 400 eV was employed. Brillouin zone sampling was carried out using a Monkhorst–Pack k-point mesh of 4 × 4 × 8. Spin polarization was not considered. The convergence criterion for the electronic self-consistent loop was set to $1 \times 10^{-5}$ eV. Structural relaxation was performed prior to mechanical property evaluation. Atomic positions were relaxed while keeping the cell shape fixed (ISIF = 0) until the residual forces on each atom were less than $1 \times 10^{-4}$ eV Å$^{-1}$. The relaxed ground-state structures were subsequently used for further analysis. The hardness was estimated solely from the optimized crystal structures using the Lyakhov–Oganov hardness model as implemented in the USPEX online[48]. The model evaluates bond strength, bond directionality, and ionicity based on the crystal bond network derived from the relaxed structures.

## 3. Results and discussion

As shown in Fig. 1(a), $Ge_{1-x}Yb_x$ thin films were deposited onto Mo foils by attaching metal chips onto a Ge sputtering target. The crystallographic state of the as-deposited $Ge_{1-x}Yb_x$ films was investigated using XRD and Raman spectroscopy. No diffraction peaks originating from crystalline Ge or $Ge_{1-x}Yb_x$ were observed in the XRD patterns, indicating that no compound formation between Ge and Yb occurred during the deposition of $Ge_{1-x}Yb_x$. Consistently, the Raman spectra exhibited a characteristic peak associated with amorphous Ge. The intensity of the amorphous Ge peak decreased with increasing Yb concentration, which can be attributed to the incorporation of Yb that disrupts Ge–Ge bonding. From these results, the $Ge_{1-x}Yb_x$ films are concluded to be in an amorphous state immediately after deposition. In addition, the Yb valence in $Ge_{0.97}Yb_{0.03}$ was qualitatively assessed using XPS measurements. In the Yb 4d spectrum, characteristic features are observed at approximately 195 eV and 185 eV. These features correspond to high- and low-binding-energy components, respectively, suggesting the coexistence of final-state contributions derived from $Yb^{2+}$ and $Yb^{3+}$ [49]. Figs. 1(b) and (c) present the charge–discharge characteristics of Ge and $Ge_{0.9}Yb_{0.1}$ anodes, respectively. Both samples exhibited stable and reproducible charge–discharge behavior over 100 cycles. The voltage plateau near 0.5 V in both samples indicates reversible lithiation/delithiation occurring within the Ge and $Ge_{0.9}Yb_{0.1}$ electrodes [6,9,12]. Notably, the $Ge_{0.9}Yb_{0.1}$ anode retained a significantly higher capacity after 100 cycles compared to the Ge anode. Figure 1d shows the Yb composition dependence of cycling performance based on discharge capacity derived from the charge–discharge characteristics. A strong dependence of cycle stability on the Yb composition was observed. For undoped Ge, the capacity rapidly declined, which can be attributed to film damage caused by the substantial volume expansion during Li insertion [16,19,20]. In contrast, incorporation levels below 10% effectively suppressed degradation, while levels above 10% resulted in a significant capacity decrease. Based on these results, the initial discharge capacity and cycle life of $Ge_{1-x}Yb_x$ anodes were evaluated and plotted as a function of Yb composition, as shown in Fig. 1(e). The initial capacity was maintained at a high level for low Yb composition (< 5%), but decreased significantly at higher composition (> 5%). This trend is reasonable, as Yb itself does not contribute to Li storage capacity. Importantly, appropriate Yb incorporation enhanced the anode lifetime by approximately a factor of three. These results demonstrate that even small additions of Yb to Ge can substantially improve cycle life without compromising the initial discharge capacity. To elucidate the origin of the improved cycle life, we focused on the rate capability and the characteristics of the formed SEI. To investigate the effect of Yb addition to Ge on the charge–discharge rate performance, rate capability measurements were conducted (Figs. 1(f), (g)). At high C-rates, both the anode capacity and the normalized anode capacity decrease upon Yb addition, suggesting that Yb incorporation degrades the rate capability of the Ge anodes. This behavior suggests a practical trade-off between enhanced cycling stability and reduced rate capability. While metal incorporation improves mechanical durability of the Ge anode, it may simultaneously introduce kinetic limitations for Li-ion transport or electronic conduction, particularly at high current densities. Notably, the undoped Ge anode retains its initial capacity beyond 30 cycles, although this cycling duration is longer than the cycling lifetime evaluated in the charge–discharge characteristics shown in Fig. 1(b). This behavior can be attributed to the lower effective capacity under high-rate cycling, thereby suppressing electrode degradation associated with Li-ion insertion/extraction. Meanwhile, a comparative XPS analysis of the SEI formed after charge–discharge cycling indicated no noticeable changes attributable to Yb addition. These results suggest that the prolonged cycle life achieved by Yb addition does not originate from an improvement in rate capability or from modifications to the SEI composition. This trade-off likely arises because larger dopants, while suppressing fracture accumulation, also reduce the continuity of electrochemically active Ge and increase local structural disorder, thereby hindering rapid Li transport under high-rate operation.

We conducted surface characterization of the $Ge_{1-x}Yb_x$ thin films after 40 charge–discharge cycles at a current density of 1 A $g^{-1}$ by disassembling the coin cells. Fig. 2 presents scanning electron microscopy (SEM) images and corresponding elemental maps of the $Ge_{1-x}Yb_x$ anodes obtained using EDX. From the SEM images of the as-deposited samples, grooves originating from the morphology of the Mo substrate were observed regardless of the $x$. This result confirms that the surface morphology does not change with varying $x$ in this range. After all charge–discharge cycles, cracks were observed in the $Ge_{1-x}Yb_x$ anodes, and partial exposure of the Mo substrate was confirmed. This behavior is attributed to the expansion and contraction of the $Ge_{1-x}Yb_x$ films during repeated lithiation and delithiation [40,41]. Post-cycling surface morphology—including the degree of cracking and delamination—exhibited a strong dependence on the Yb composition. In particular, the undoped Ge sample ($x$ = 0%) showed more severe delamination compared to the Yb-doped counterparts. This trend is consistent with the electrochemical results shown in Fig. 1, where Yb incorporation was found to enhance the cycling stability of the Ge anodes. As the Yb concentration increased, the morphology of the Ge–Yb islands gradually evolved into more rectangular shapes, which reflected the polishing direction of the underlying Mo substrate. This observation suggests that Yb incorporation alters the mechanical response of the Ge layer to lithiation-induced stress, potentially influencing how structural damage develops within the anode during cycling.

Figs. 3(a)-(j) presents cross-sectional TEM and EDX images of the Ge and $Ge_{0.97}Yb_{0.03}$ anodes after electrochemical cycling. Due to severe delamination observed in the Ge sample after 40 cycles, samples after 10 cycles were used for comparison. To facilitate FIB processing,

the sample surfaces were coated with a thin Pt layer. From the bright-field TEM, high-angle annular dark field scanning TEM (HAADF-STEM) images, and EDX elemental maps, the Ge(Yb) layer, Mo substrate, and interface region were clearly distinguishable in both samples. In the undoped Ge sample, the Ge islands exhibited significant curvature and detachment at the Ge/Mo interface, indicating mechanical degradation. In contrast, the $Ge_{0.97}Yb_{0.03}$ sample showed a uniform nanometer-scale distribution of Yb within the Ge matrix, and the extent of curvature and delamination was substantially reduced compared to the undoped sample. These observations suggest that Yb incorporation effectively suppresses the deformation and interfacial delamination of the Ge layer, thereby contributing to the improved electrochemical performance of the anode. Cross-sectional SEM observations were performed to examine whether the volumetric expansion during lithiation differs depending on Yb incorporation. The as-deposited film thicknesses of Ge and $Ge_{0.97}Yb_{0.03}$ were both 500 nm; however, after lithiation, both films were found to expand to approximately 1.8 μm (Figs. 3(k) and (l)). Accordingly, the volumetric expansion ratio is estimated to be approximately 260%, which is in good agreement with previously reported results [50,51].

We investigated the effect of different metal species on the electrochemical performance of Ge anodes. Here, metal elements were selected to span a wide range of atomic sizes, and the amount of the added metal was fixed at 3%, as this corresponds to a small addition level that allows relatively easy compositional control while remaining low enough that conventional mitigation of lithiation-induced volumetric expansion does not occur, as already revealed in Fig. 3(k) and (l). Here, an improvement in cycle life was also observed even with 0.6% Yb addition, indicating that 3% is by no means a singular or exceptional point. Prior to electrochemical measurements, the crystallographic state of the $Ge_{0.97}M_{0.03}$ films was examined using XRD and Raman spectroscopy. The XRD patterns showed no diffraction peaks originating from crystalline Ge or Ge-related compounds, and only peaks from the substrate were observed. This result indicates that, similar to $Ge_{0.97}Yb_{0.03}$, no compound formation between Ge and M occurred during the deposition of $Ge_{0.97}M_{0.03}$. Consistently, Raman measurements revealed characteristic peaks associated with amorphous Ge for all metal species. From these results, the $Ge_{0.97}M_{0.03}$ films are confirmed to be amorphous. For the Ge–Yb system, cross-sectional TEM and EDX mapping confirm a spatially uniform distribution of Yb without detectable segregation within the experimental resolution, supporting atomic-scale incorporation. For the other metal species, although XRD and Raman measurements indicate the absence of crystalline secondary phases, detailed nanoscale chemical mapping comparable to that performed for Yb was not conducted. Therefore, the possibility of nanoscale clustering or short-range compositional inhomogeneity cannot be completely excluded. The mechanistic interpretation presented here should therefore be understood within this structural scope. Under these structural conditions, clear charge–discharge behavior was observed for all samples, indicating that incorporation did not hinder Li insertion/extraction. As shown in Fig. 4a, the cycle performance of the anodes exhibited a strong dependence on the type of incorporated metal. Focusing on the atomic size of the incorporated metals, we examined the correlation between atomic size and anode performance. Figure 4b displays the initial discharge capacity and cycle life of $Ge_{0.97}M_{0.03}$ anodes as a function of incorporated metal atomic size. To quantitatively evaluate the relationship between the incorporated metal atomic size and the electrochemical performance, Pearson's correlation coefficients were calculated [52]. The correlation coefficients for the initial discharge capacity ($r_{CA}$) and the cycle life ($r_{LA}$) were defined as

$$r = \frac{\sum_{i=1}^{n}(x_i-\bar{x})(y_i-\bar{y})}{\sqrt{\sum_{i=1}^{n}(x_i-\bar{x})^2}\sqrt{\sum_{i=1}^{n}(y_i-\bar{y})^2}}\ , \qquad (1)$$

where $x_i$ represents the atomic radius of the incorporated metal for the $i$-th sample, $y_i$ corresponds to either the initial discharge capacity (for $r_{CA}$) or the cycle life (for $r_{LA}$), and $\bar{x}$ and $\bar{y}$ denote their respective mean values. The number of data points $n$ corresponds to the number of incorporated elements investigated (Ge, Al, Cu, Ni, Ag, W, Ta, and Yb). The atomic sizes of the incorporated elements were taken as atomic radii reported in the literature [53], while the electrochemical parameters (initial discharge capacity and cycle life) were directly extracted from the experimental charge–discharge measurements shown in Fig. 4(a). The initial capacity remained nearly constant at approximately 1300 mAh $g^{-1}$, regardless of incorporated element size, with a weak correlation coefficient ($r_{CA} = -0.5$) between atomic size and capacity. In contrast, the cycle life showed a clear increasing trend with incorporated element's atomic size, exhibiting a remarkably strong positive correlation ($r_{LA} = 0.95$). These results demonstrate that the longevity of Ge anodes is strongly correlated with the atomic size of the incorporated element, with Yb yielding the longest cycle life among all tested elements. For convenience, the electrochemical properties obtained in this section are also summarized in Table S1.

We conducted nanoindentation measurements to investigate the origin of the improved performance of Ge anodes achieved by metal incorporation. In the nanoindentation test, as shown in Fig. 5(a), a sharp tip was pressed into the film to simultaneously evaluate the Young's modulus and hardness of the thin film [54]. After the test, an indentation remained on the film surface, as shown in Fig. 5(b). From the indentation region, the hardness ($H$) and Young's modulus of $Ge_{0.97}M_{0.03}$ ($E_{\text{sample}}$) were calculated using the following equations:

$$H = \frac{P}{A} \tag{2}$$

$$S = \frac{dP}{dh} \mid h = h_{\text{max}} \tag{3}$$

$$E_{\text{r}} = \frac{S\sqrt{\pi}}{2\sqrt{A}} \tag{4}$$

$$\frac{1}{E_{\text{r}}} = \frac{1-\nu_{\text{sample}}^2}{E_{\text{sample}}} + \frac{1-\nu_{\text{indenter}}^2}{E_{\text{indenter}}} \tag{5}$$

Here, $P$ is the applied load, $A$ is the projected contact area, $h_{\text{max}}$ is the maximum indentation depth, $E_{\text{r}}$ is the reduced modulus that accounts for the combined elasticity of the diamond indenter and the $Ge_{0.97}M_{0.03}$ sample, and $\nu_{sample}$ *and* $\nu_{indenter}$ are the Poisson's ratio for the sample and indenter. We used the following elastic constants: $E_{indenter}$ = 1141 GPa and $\nu_{indenter}$ = 0.06 for the diamond indenter, and $\nu_{sample}$ = 0.28 for $Ge_{0.97}M_{0.03}$. The $h_{max}$ value was set to 50 nm, approximately one-tenth of the film thickness, to minimize the influence of the substrate. For the measurements, $Ge_{0.97}M_{0.03}$ films were deposited on flat glass substrates, and 15 measurements were conducted for each sample. We selected four representative metals (Ge, Ni, Ta, Yb) with varying atomic sizes and measured their Young's modulus and hardness. Figures 5c and 5d show the Young's modulus and hardness of the $Ge_{0.97}M_{0.03}$ films, along with their correlation coefficients with atomic size. The correlation coefficient between Young's modulus and atomic size was $r_{YA}$ = −0.15, indicating little correlation. Conversely, the correlation coefficient between hardness and atomic size was $r_{HA}$ = −0.93, showing a clear trend of decreasing hardness with increasing atomic size. This behavior can be interpreted as a solid-solution softening effect caused by the addition of heteroatoms [55]. Moreover, the correlation coefficient between hardness and anode cycle life, $r_{LH}$, was also −0.86, clearly demonstrating that softening of the anode hardness leads to improved cycle life. It should be noted that the hardness measured for the metal-incorporated Ge films in this study is a surrogate metric and does not directly quantify fracture toughness during Li insertion and extraction. Nevertheless, we consider hardness to be a relevant correlating parameter for cycle-life improvement, as supported by the systematic trends observed across different incorporated metals. Regarding the physical mechanism underlying the prolonged cycle life associated with mechanical softening, we currently interpret that reduced hardness facilitates the formation of cracks in the Ge anode during repeated Li-ion insertion and extraction. This crack formation leads to fragmentation of the film into finer islands, which in turn makes the Ge film less prone to large-scale delamination from the substrate after extended cycling as shown in Fig. 2. As a result, electrical connectivity and structural integrity are better preserved over long-term cycling. We emphasize that this interpretation represents a physically motivated inference based on the observed trends, rather than a direct experimental determination of the damage evolution process. To assess whether metal incorporation into Ge thin films can reasonably lead to a reduction in hardness, first-principles calculations were performed. The $H_v$ variation was evaluated by applying strain to a supercell structure in which one of the 32 Ge atoms was substituted with a metal element (Fig. 5(e)). The calculations clearly reveal a trend in which the $H_v$ decreases with increasing atomic size of the metal (Fig. 5(f)). This tendency is consistent with the experimental results shown in Fig. 5(d), suggesting that the observed reduction in hardness originates from local structural distortion induced by the incorporation of different metal elements. This interpretation is in line with the general understanding of amorphous alloy systems, in which atomic-size mismatch disrupts short-range order, reduces structural coherence, and thereby lowers the effective rigidity of the amorphous Ge network [56]. Although the present study establishes an empirical correlation between dopant atomic size, mechanical softening, and electrochemical lifetime, the framework remains primarily descriptive. Machine learning approaches could incorporate additional material descriptors and operating parameters to enable robust prediction under uncertainty and stochastic degradation conditions [57].

## 4. Conclusion

This study introduces a new paradigm in electrode material design, namely, mechanical softening via metal incorporation, to overcome the intrinsic limitations of Ge-based LIB anodes. Unlike most previous strategies that rely on structural design concepts, such as nanostructuring, composite formation, or volume buffering, the present approach is materials-intrinsic and aims to improve damage tolerance by reducing the hardness of the active material itself. By incorporating Ge with small amounts (<3%) of various metal elements, we demonstrate a substantial enhancement in cycle life that systematically increases with the atomic size of the incorporated metals without sacrificing initial capacity. Through comprehensive structural and electrochemical characterization, we attribute this improvement to a reduction in film hardness, which effectively suppresses lithiation-induced mechanical damage such as cracking and delamination. The observed correlation between incorporated metal's atomic size and mechanical softening provides a useful framework for understanding and optimizing the mechanical response of amorphous Ge-based anodes. Although the present discussion is limited to thin-film systems, the use of thin-film model electrodes enables a controlled investigation of intrinsic mechanical–electrochemical coupling by minimizing extrinsic effects such as binder chemistry, particle–particle interfaces, and heterogeneous current distribution. The translation of the present concept to practical electrodes will require extension to particulate or composite architectures, optimization of binder and conductive additive systems capable of accommodating controlled cracking, and validation at higher areal loadings. Nevertheless, the underlying concept of mechanically softening high-capacity alloy anodes through atomic-scale incorporation may be broadly applicable to other Ge-based electrode architectures, including bulk, particulate or composite systems, although further studies will be required to validate its effectiveness under practical electrode conditions. At the same time, however, these findings provide a new and compelling perspective on the mechanical design of high-capacity electrode materials and hold significant promise for advancing the frontiers of energy storage technologies, spanning applications from portable electronics to electric vehicles and grid-scale energy storage systems.

## ACKNOWLEDGEMENTS

This study was financially supported by the KUMA Foundation, JSPS KAKENHI (No. 22K18802), and JST FOREST Program (No. JPMJFR222J). Certain experiments were conducted at the Advanced Research Infrastructure for Materials and Nanotechnology in Japan.

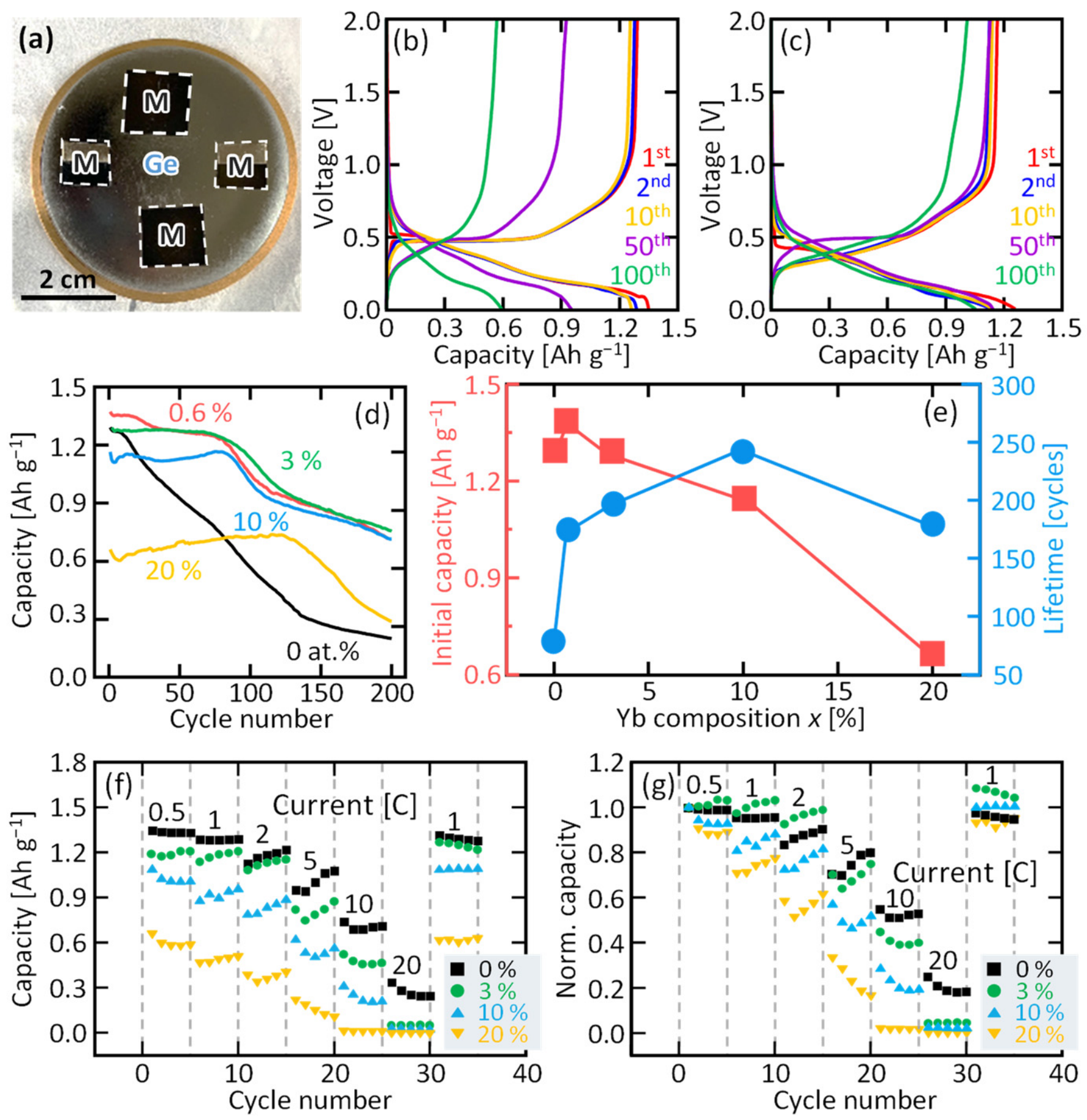

**Figure 1.** Effects of Yb addition on Ge anodes. (a) Photograph of the Ge sputtering target with Yb chips affixed to its surface. Electrochemical characterization of the $Ge_{1-x}Yb_x$ anodes in a coin-type cell at a current density of 0.5 A $g^{-1}$: charge-discharge profiles of the (b) Ge and (c) $Ge_{0.9}Yb_{0.1}$ anodes, (d) discharge capacity of the $Ge_{1-x}Yb_x$ anodes as a function of cycle number, and (e) initial discharge capacity and lifetime as functions of $x$. (f, g) Rate capability of $Ge_{1-x}Yb_x$ anodes at different C-rates (0.5–20 C): (f) discharge capacity and (g) normalized discharge capacity. Here, the capacity was normalized to the first-cycle capacity.

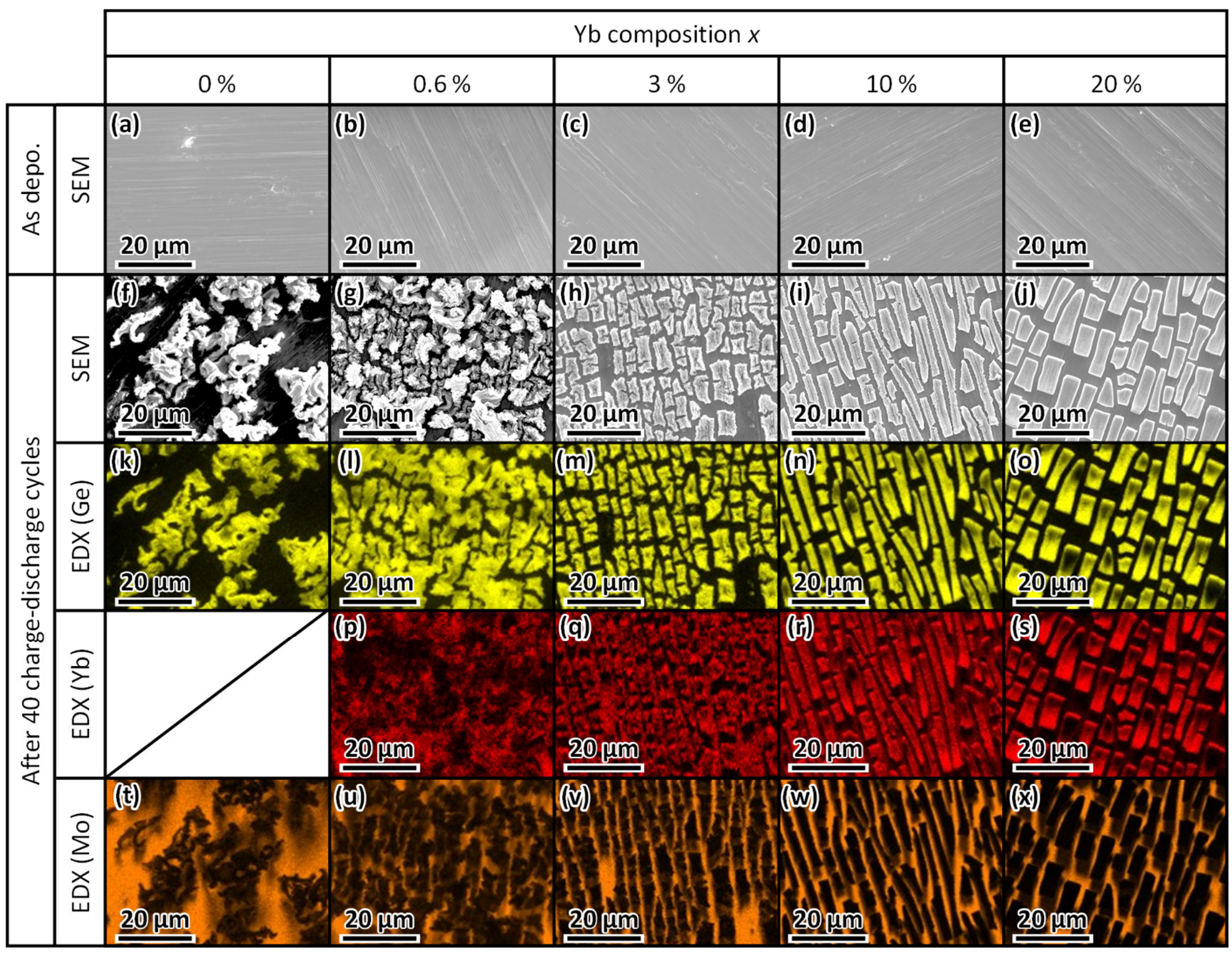

**Figure 2.** Surface characterization of $Ge_{1-x}Yb_x$ anodes ($x$ = 0, 0.6, 3, 10, and 20%). SEM images of the anodes (a–e) before and (f–j) after 40 charge–discharge cycles at a current density of 1 A $g^{-1}$. Corresponding EDX elemental maps after cycling showing the distributions of (k–o) Ge, (p–s) Yb, and (t–x) Mo.

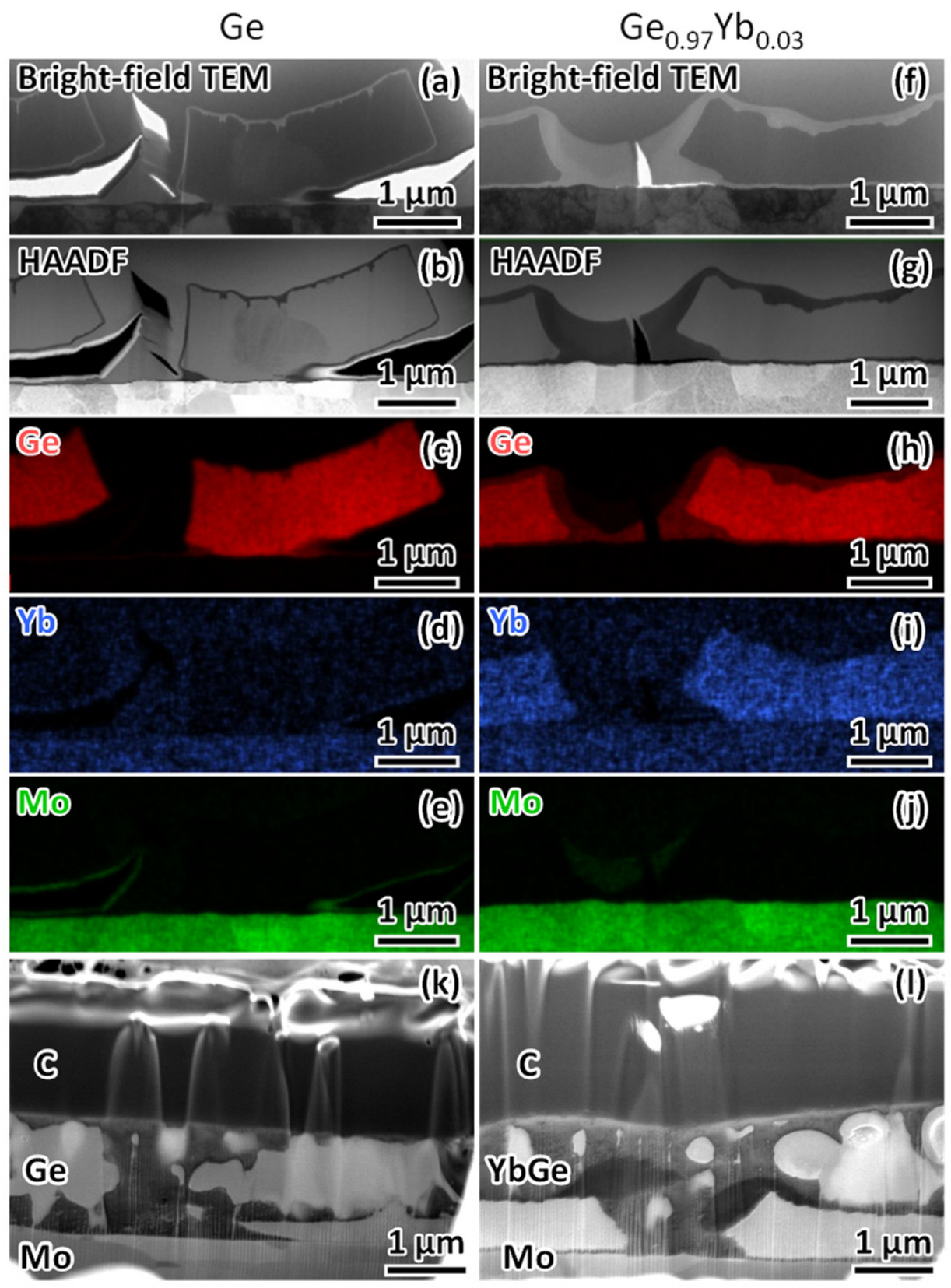

**Figure 3.** (a–j) Cross-sectional TEM analysis of the (a–e) Ge and (f–j) $Ge_{0.97}Yb_{0.03}$ anodes on a Mo substrate after 10 charge–discharge cycles at a current density of 1 A $g^{-1}$. (a,f) Bright-field TEM images. (b,g) HAADF-STEM images. EDX elemental maps of (c,h) Ge, (d,i) Yb, and (e,j) Mo. (k, l) Cross-sectional SEM images of the anodes in the charged state during the second cycle for (k) Ge and (l) $Ge_{0.97}Yb_{0.03}$. A thin carbon layer was deposited on the sample surfaces as a protective layer prior to FIB preparation and electron microscopy observations.

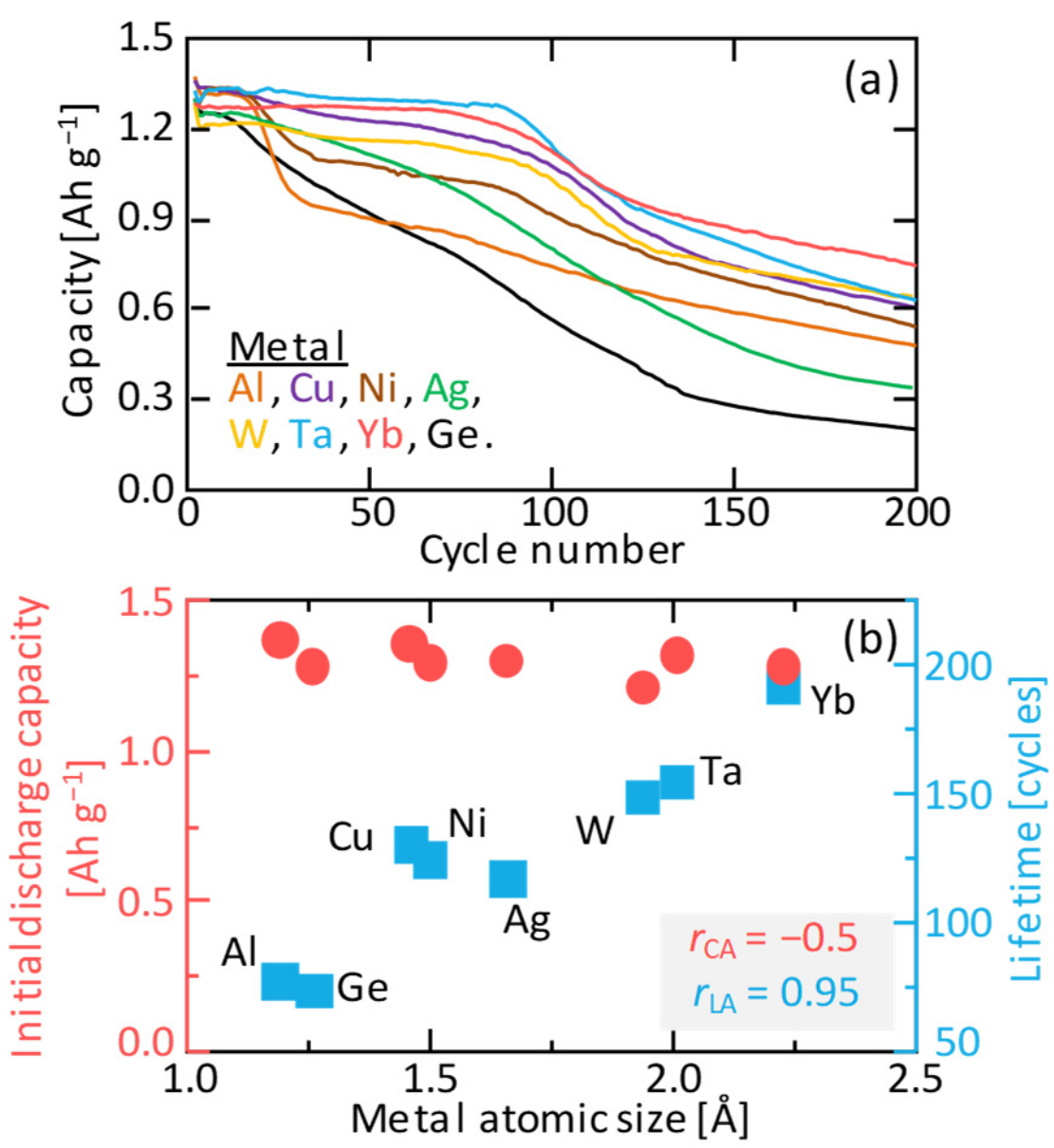

**Figure 4.** Electrochemical performance of the $Ge_{0.97}M_{0.03}$ (M = Al, Cu, Ni, Ag, W, Ta, Yb, and Ge) anodes at a current density of 0.5 A g$^{-1}$. (a) Discharge capacity as a function of cycle number. (b) Initial discharge capacity and lifetime as functions of the atomic size of metals. $r_{LA}$ and $r_{CA}$ denote the correlation coefficients between atomic size of metals and the lifetime and initial discharge capacity of the $Ge_{0.97}M_{0.03}$ anodes, respectively.

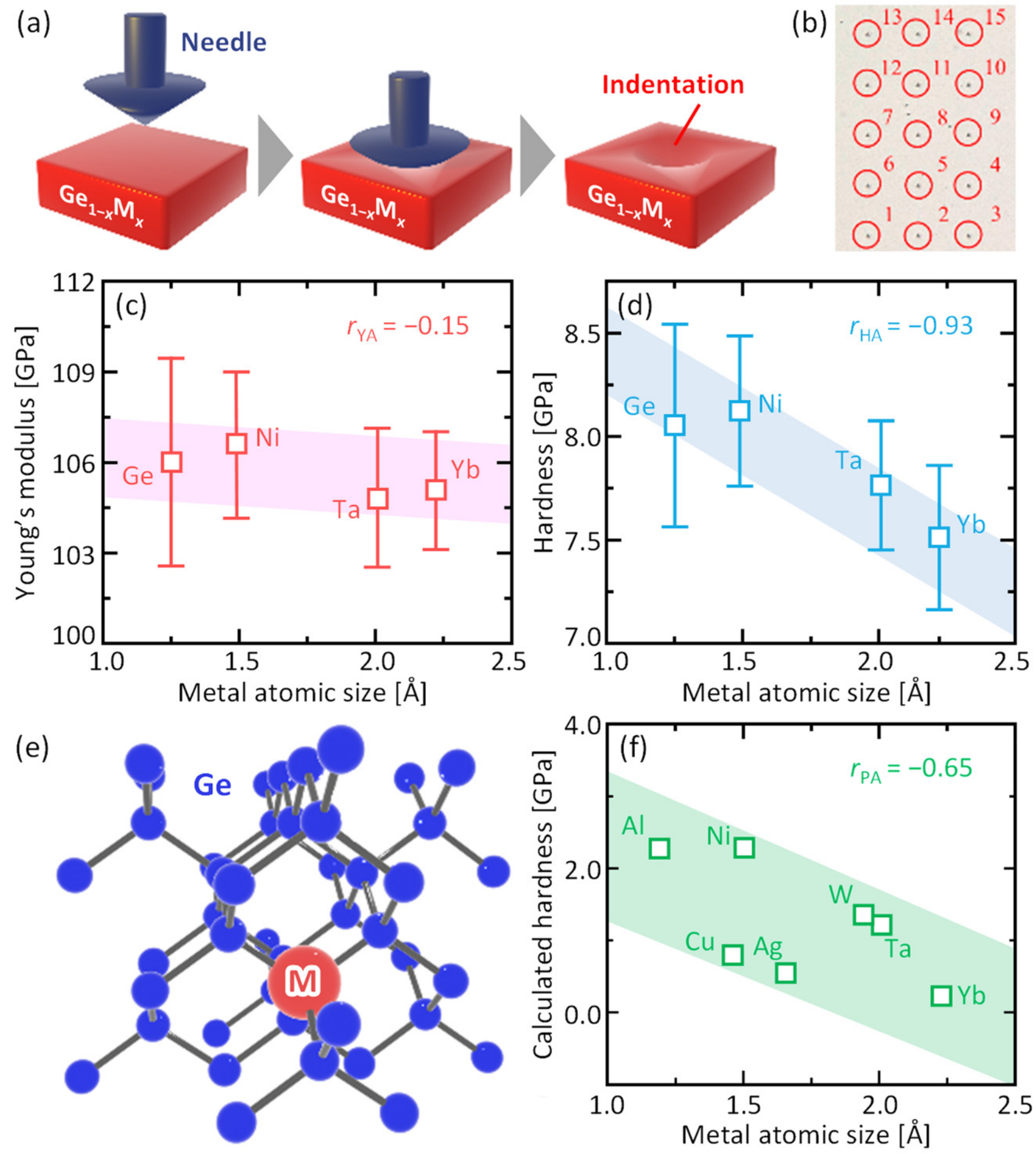

**Figure 5.** Mechanical properties of the $Ge_{0.97}M_{0.03}$ anodes evaluated using (a)–(d) nanoindentation measurements and (e), (f) calculations. (a) Schematic illustration of the nanoindentation setup. (b) Optical micrograph showing the surface of the $Ge_{0.97}Yb_{0.03}$ sample after indentation. (c) Young's modulus and (d) hardness of the $Ge_{0.97}M_{0.03}$ anodes as functions of metal atomic size. Nanoindentation measurements were conducted on samples prepared on $SiO_2$ substrates. $r_{YA}$ and $r_{HA}$ denote the correlation coefficients between metal atomic size and the Young's modulus and hardness of the $Ge_{0.97}M_{0.03}$ anodes, respectively. (e) Schematic illustration of the Ge-based supercell in which one Ge atom is substituted by a metal atom, corresponding to an incorporated element's concentration of approximately 3 %. (f) Hardness values estimated from the calculated elastic constants using the empirical model proposed by Lyakhov *et al.* [48]. $r_{PA}$ denote the correlation coefficients between metal atomic size and the calculated hardness.